\documentclass[preprint]{jfm}
\usepackage{graphicx}

\usepackage{color}
\usepackage{siunitx}

\title{Role of natural convection in the dissolution of sessile droplets}
\author{Erik Dietrich \aff{1,2},
Sander Wildeman \aff{1},
Claas Willem Visser \aff{1},
Kevin Hofhuis \aff{2},
E. Stefan Kooij \aff{2},
Harold J. W. Zandvliet \aff{2},
\and
Detlef Lohse \aff{1}\corresp{\email{d.lohse@utwente.nl}} }
\affiliation{\aff{1}Physics of Fluids, 
\aff{2}Physics of Interfaces and Nanomaterials, MESA+ Institute for Nanotechnology, University of Twente, P.O. Box 217, 7500 AE Enschede, The Netherlands.}
\begin{document}
\maketitle
\def\mean#1{\left< #1 \right>}
\begin{abstract}
The dissolution process of small (initial (equivalent) radius $R_{0}<1$ mm) long-chain alcohol (of various types) sessile droplets in water is studied, disentangling diffusive and convective contributions. The latter can arise for high solubilities of the alcohol, as the density of the alcohol-water mixture is then considerably less as that of pure water, giving rise to buoyancy driven convection. The convective flow around the droplets is measured, using micro-particle image velocimetry (\si{\micro}PIV) and the schlieren technique. When non-dimensionalizing the system, we find a universal $\text{Sh} \sim \text{Ra}^{1/4}$ scaling relation for all alcohols (of different solubilities) and all droplets in the convective regime. Here $\text{Sh}$ is the Sherwood number (dimensionless mass flux) and $\text{Ra}$ the Rayleigh number (dimensionless density difference between clean and alcohol-saturated water). 
This scaling implies the scaling relation $\tau_{\text{c}}\propto R_0^{5/4}$ of the convective dissolution time $\tau_{\text{c}}$, which is found to agree with experimental data. We show that in the convective regime the plume Reynolds number (the dimensionless velocity) of the detaching alcohol-saturated plume follows $\text{Re}_\text{p}\sim \text{Sc}^{-1}\text{Ra}^{5/8}$, which is confirmed by the \si{\micro}PIV data. Here, $\text{Sc}$ is the Schmidt number. The convective regime exists when $\text{Ra}>\text{Ra}_{\text{t}}$, where $\text{Ra}_{\text{t}}=12$ is the transition Ra-number as extracted from the data. For $\text{Ra}\leq\text{Ra}_{\text{t}}$ and smaller, convective transport is progressively overtaken by diffusion and the above scaling relations break down.
\end{abstract}


\section{Introduction}
Conventional wisdom says that oil and water do not mix. However, some oily liquids, e.g. long-chain alcohols, are slightly soluble in water (see table 1). When a droplet of such an alcohol is placed in a bath of water, it will slowly dissolve. Figure 1 shows an example of a sessile 1-hexanol droplet in water for which the dissolution time $\tau$ was about 3 hours. Considering this long dissolution time, it may seem plausible to assume that mass transport away from the droplet is governed by diffusion. Equivalent to the diffusion driven mass transport from small gas bubbles \citep{Epstein1950} or small sessile droplets \citep{Popov2005, Stauber2014, Zhang2015, Lohse2015} the relevant time-scale would in this case then be given by 
\begin{equation}
\tau_{\text{d}} =\frac{ R_0^2 \rho_{\text{d}}}{2 D\Delta c}
\label{difftime}
\end{equation}
where $R_0$ is the initial equivalent radius of the droplet, $D$ the diffusion constant of the alcohol in water, $\rho_{\text{d}}$ is the density of the droplet material, and $\Delta c$ is the difference between the saturated concentration $c_s$ at the droplet interface and the (undersaturated) concentration $c_{\infty}<c_s$ far away from the drop. However, for the 1-hexanol droplet with an initial radius $R_{0}=0.7$ mm, one finds $\tau_{\text{d}} \approx 11$ hours, which is much longer than the 3 hours observed experimentally. In previous work \citep{Dietrich2015} we hypothesized that this discrepancy is caused by the neglect of buoyancy driven convection of the slightly lighter alcohol-water mixture near the droplet interface. The same idea has been put forward in the context of slowly growing CO$_2$ bubbles in small supersaturations \citep{Enriquez2014, Penas2015}, and evaporating droplets \citep{Shahidzadeh2006}. Also in these cases, the rate of mass transport exceeded the diffusion limited predictions. On the other hand, even for mm-sized droplets, there also seem to be circumstances under which the diffusive time scale is accurate \citep{Picknett1977, Gelderblom2011, Stauber2014, Stauber2015a}. This reflects the existence of a threshold for convection. However, the details of this threshold remain unclear. The low velocities and small refractive index differences involved often inhibit direct observation of the buoyant flow in the surrounding medium. As far as we are aware, direct visualization attempts of the external flow were only undertaken in the context of evaporating droplets, using either schlieren \citep{KellyZion2013}, infrared spectroscopy \citep{kelly2013vapor}, interferometry \citep{Dehaeck2014}, and only very recently, by tracing tiny oil droplets in air \citep{Somasundaram2015}.

In this work we combine qualitative schlieren imaging with quantitative micro-particle image velocimetry (\si{\micro}PIV) to directly visualize the concentration field and flow around slowly dissolving droplets of various types of long-chain alcohols in clean water. We show that above a transition solutal Rayleigh number, which corresponds to the buoyancy of the alcohol-water mixture, which is lighter than the surrounding clean water, the solute is mainly transported away in a single steady plume above the droplet. Knowledge of the flow structure allows us to derive scaling laws for both the dissolution rate and the plume velocity in the convective regime, which are in good agreement with experiments. Finally, as the droplet shrinks and its Rayleigh number drops below the transition value, a transition occurs in which convection dies out and is overtaken by diffusion.

\begin{figure}
\begin{center}
\includegraphics[angle=0,width=12cm]{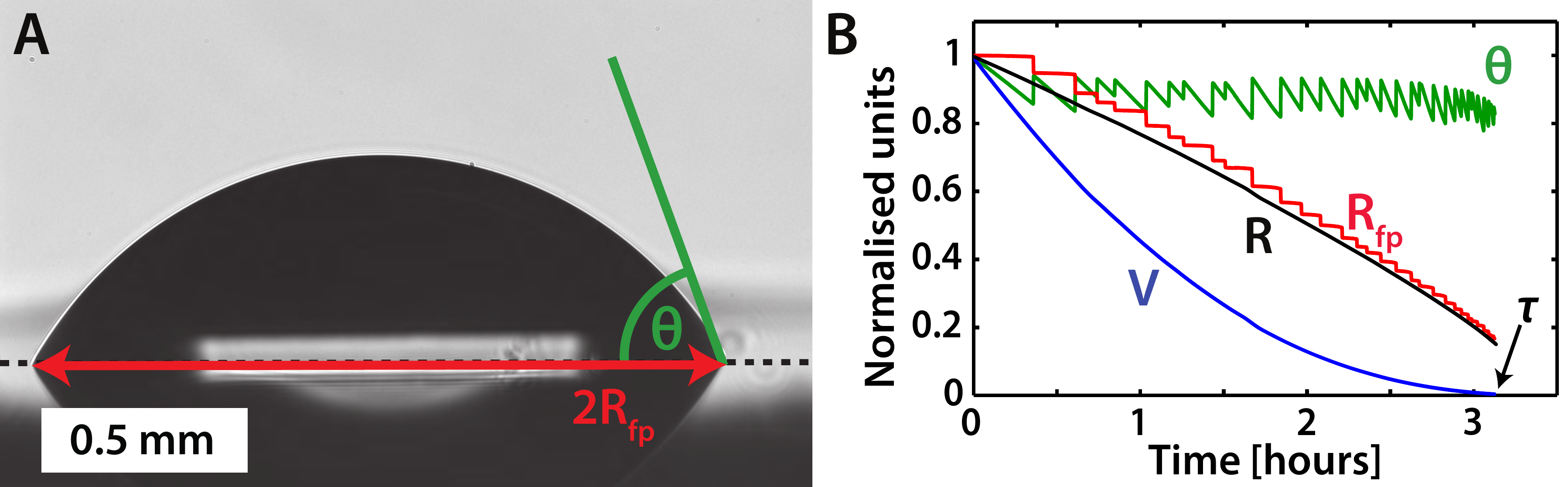}
\end{center}
\caption{\label{figure1} (A) Snapshot of a dissolving long-chain alcohol droplet (1-hexanol) in water. The dotted black line indicates the location of the silicon substrate, which mirrors part of the droplet. The footprint radius $R_{\rm{\rm{fp}}}$ and the contact angle $\theta$ are indicated. A movie of the entire dissolution process is available online as supplementary material. (B) Evolution of the aforementioned parameters in time, along with the volume $V$ of the droplet and the equivalent radius $R=\left(3V/2\pi\right)^{1/3}$. All parameters have been normalised by their initial values: $V_0=750$ nL, $R_{\rm{fp},0}=0.825$ \si{\milli\metre}, $R_{0}=0.708$ \si{\milli\metre}, and $\theta_0=72^{\circ}$.}
\end{figure}


\section{Experimental Procedure}
\subsection{Materials and Preparation}
As shown in table \ref{tab:matprop} the solubility of long-chain alcohols strongly depends on their length, while other properties, like density and diffusion coefficient are relatively insensitive to this. By increasing the number of carbon atoms in the chain from 5 (pentanol) to 8 (octanol), one decreases the solubility (and thereby the buoyant force of the water-alcohol mixture) by two orders of magnitude. This makes these alcohols very suitable to study the possible transition between diffusion and convection.
Alcohols with purities of $\geq98\%$ (Sigma-Aldrich) were used. The density of the alcohol-water mixture (table \ref{tab:matprop}) was calculated for a mixture at $100\%$ saturation, using the molal volume $\phi_V^0$ at infinite dilution \citep{Hoiland1976, Romero2007}. The molal volume $\phi_V$ gives the volume occupied by one mole of solute in the solvent. The assumption is made that $\phi_V$ is independent of the solute concentration, which introduces a negligible error in $\Delta\rho$ of $<1\%$ when compared to the direct density measurements given by \citet{Romero2007}. 

\begin{table}
\scriptsize
\caption{Properties of the alcohols used in this work, measured at $25^{\circ}$C: Chemical composition, density $\rho$ of the pure alcohol, diffusion constant $D$ of the alcohol in water, saturation solubility $c_s$ of the alcohol in water, molal volume $\phi_V^0$ of the alcohol in water at infinite dilution, density difference $\Delta\rho=\rho_{H_2O}-\rho_{H_2O,\text{sat}}$ between clean water and water saturated with the alcohol, interfacial tension $\gamma$ of the water-alcohol interface. The diffusive timescale $\tau_{\text{d}}$ is calculated according to equation (\ref{difftime}), for $R_0=0.7$ mm. No values for $\gamma$ for 2-heptanol and 3-heptanol could be found. Data were obtained from \citet{Crittenden1954}$^{\ddagger}$, \citet{Kinoshita1958}$^{\ast}$, \citet{Hoiland1976}$^{\spadesuit}$, \citet{Stephenson1984}$^{\flat}$, \citet{Demond1993}$^{\clubsuit}$, \citet{Hao1996}$^{\dagger}$, \citet{Romero2007}$^{\star}$, and \citet{yalkowsky2010handbook}$^{\blacklozenge}$. Diffusion constants for 2-heptanol and 3-heptanol were assumed to be equal to the 1-heptanol value, and the diffusion constant of 1-octanol was obtained by extrapolating data from \citet{Hao1996}.}
\begin{tabular}{ |c | c | c | c | c |c|c|c|c|}
\hline
Alcohol 	&  Composition			& $\rho$			& D 											&   $c_s$ 							&$\phi_V^0$ 						&$\Delta\rho	$			&  $\gamma$		& $\tau_{\text{d}}$\\
		&					& [kg m$^{-3}$]		& [$10^{-9}$m$^2$s$^{-1}$]							&[kg m$^{-3}$]						&[cm$^3$mol$^{-1}$]				&[kg m$^{-3}$]			&[mN/m]			&[$10^4$ s]		\\
\hline
1-Pentanol 	&	C$_5$H$_{11}$OH		&   $811$			&  $0.888^{\dagger}$ 				 				&  $22^{\ddagger\ast}$ 					&$102.62^{\star\spadesuit}$ 			& $3.42$				& $4.4^{\clubsuit}$	&$1.0$	\\
1-Hexanol 	&  	C$_6$H$_{13}$OH		&   $814$ 		 	&  $0.83^{\dagger}$ 								&  $5.9^{\ddagger\ast}$					&$118.65^{\star\spadesuit}$ 			& $0.92$		 		&$6.8^{\clubsuit}$		& $4.0$   \\
1-Heptanol	&	C$_7$H$_{15}$OH		&  $822$			&  $0.800^{\dagger}$ 								&  $1.67^{\ast}$						&$136.95^{\star}$ 				& $0.29$ 				&$7.7^{\clubsuit}$		&$14.5$	\\
2-Heptanol	&	C$_7$H$_{15}$OH		&  $817$			&  $0.800$ 										&  $3.5^{\flat}$ 						&$134.39^{\spadesuit}$				&$0.53$				&-				&$7.6$		\\
3-Heptanol	&	C$_7$H$_{15}$OH		&  $818$			&  $0.800$ 										&  $4.0^{\ddagger\blacklozenge}$			&$133.30^{\spadesuit}$				&$0.57$				&-				&$5.8$		\\
1-Octanol	&	C$_8$H$_{17}$OH		&  $827$			&  $0.780$										& $0.5^{\ddagger\ast}$ 					&$148.41^{\star}$ 				&$0.07$				&$8.52^{\clubsuit}$	&$48.1$	\\
    \hline
  \end{tabular}
\label{tab:matprop}
\end{table}

\begin{figure}
\begin{center}
\includegraphics[angle=0,width=12cm]{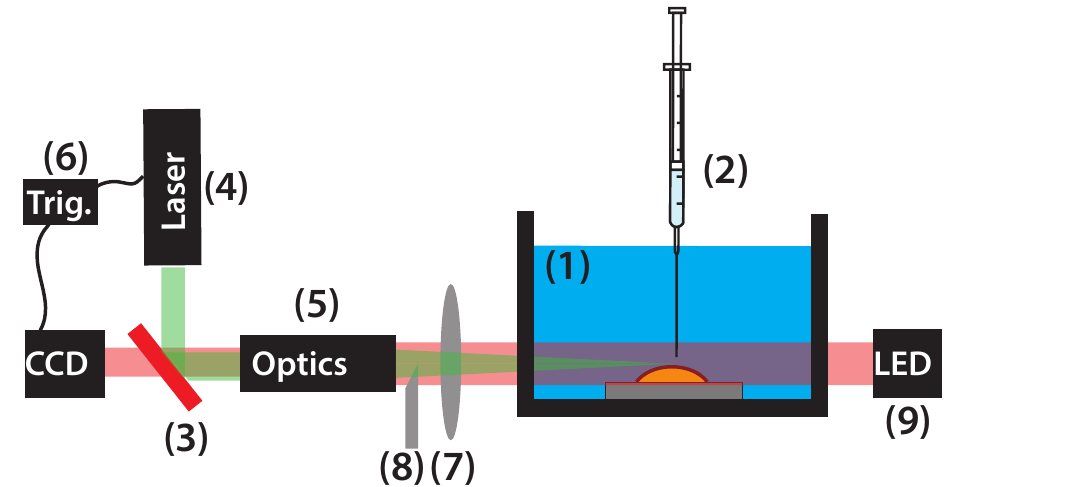}
\end{center}
\caption{\label{figure2} (color online) Experimental setup, showing the glass tank (1), with the substrate and the droplet in place (not to scale). The droplet was deposited under water using a syringe (2) fitted in a motorized syringe pump (not drawn). A dichroic mirror (3) was used to couple the laser beam (4) into the long distance microscope (5). A trigger-delay box (6) synchronized the laser pulses with the camera exposure. Parts (3), (4), and (6) were used in the \si{\micro}PIV measurements only. The assembly used for schlieren, consisting of a positive lens (7) and a knife edge (8) located in the focal point of the lens. The parallel LED light source (9) was used in all experiments, with the exception of the PIV measurements.}
\end{figure}

A sketch of the experimental setup is provided in figure \ref{figure2}. All measurements were conducted in a qubic glass tank of $5\times5\times5$ cm$^3$. The container was cleaned using isopropyl-alcohol and water, and then filled with $100$ \si{\milli\litre} of clean water. This water was obtained from a Reference A+ system (Merck Millipore, at $18.2$ \si{\mega\ohm\centi\metre}) several hours before the measurement and stored in a clean flask to equilibrate and thus reduce thermal convective currents. After the tank was filled, a single droplet was dispensed from a glass syringe with a teflon plunger, fitted in a motorized syringe pump. The droplet was placed on a hydrophobized silicon wafer $\approx 1\times1$ cm$^2$ (P/Boron/(100), Okmetic), placed at the bottom of the tank. Hydrophobization was achieved by coating the wafer with a self-assembled monolayer of PFDTS (1H,1H,2H,2H-Perfluorodecyltrichlorosilane 97\%, ABCR Gmbh, Karlsruhe Germany), following the procedure described earlier \citep{Karpitschkathesis}. Prior to each experiment the samples were cleaned by insonication in acetone for $10$ \si{\minute} and dried under a stream of nitrogen. After the droplet was placed on the substrate, the needle was removed and the tank was closed. 


\subsection{Imaging}
The droplet was illuminated from one side using a collimated LED light source (Thorlabs, wavelength $\lambda=625$ nm) and imaged onto a CCD camera (Pixelfly USB, PCO Germany) with a long-distance microscope providing a magnification up to $16\times$. The images were recorded at a rate of 1 frame per second, and post-processed using a Matlab code to extract the droplet profile with sub-pixel accuracy \citep{vandermeulen2014}. Since all droplets were smaller than the capillary length ($\sqrt{\gamma/ (\rho_{H_2O}-\rho_{alcohol}) g} \approx2$ mm), the droplet profile could accurately be fitted to a spherical cap to obtain the radius of curvature and contact angle.
With this method, droplets could be traced until $V<0.05V_0$.

\subsection{Schlieren}
The concentration gradients developing around the dissolving droplets were qualitatively visualized using the schlieren technique \citep{Settles2001}. For this, a positive lens and a knife edge were placed at the camera side of the tank, as shown in figure \ref{figure2}. After passing through the tank, the parallel light from the LED source is focused onto the edge of a sharp knife placed perpendicular to the beam. To be sensitive to both horizontal and vertical concentration gradients the knife edge was placed under an angle of $45^{\circ}$. In the resulting image, solute rich regions are visible as a local changes in light intensity.

\subsection{$\mu$PIV measurements}
For the \si{\micro}PIV measurements the water in the tank was seeded with red-fluorescent tracer particles (Fluoro-Max, Thermo Fisher Scientific, $3$ \si{\micro\metre} diameter). A pulsed green laser, (Nd:YAG, $\lambda=532$ nm) was coupled into the microscope by a dichroic mirror. The focal plane of the microscope was centered at the droplet. Therefore, only tracer particles in the $\approx100$ \si{\micro\metre} thick focal plane were imaged, producing a 2-dimensional velocity field around the symmetry axis of the droplet. The red light ($\lambda=612$ \si{\nano\metre}) emitted by the fluorescent particles was recorded by the CCD camera at $8$ frames per second. A BNC 575 pulse/delay generator was used to synchronize the laser pulse and the camera exposure. 
The obtained images were then post-processed in ImageJ to remove static features and to enhance the contrast. Consecutive image pairs were analyzed with JPIV, using an interrogation window of $32\times32$ pixels, corresponding to $\approx70\times70$ \si{\micro\metre\squared}. The concentration of tracer particles was kept low to avoid excessive absorption and blurring by out of focus particles. Because of this low particle density, the velocity fields from multiple images pairs were combined for improved accuracy \citep{raffel2007}. 

\section{Visualization results}

\begin{figure}
\begin{center}
\includegraphics[angle=0,width=\textwidth]{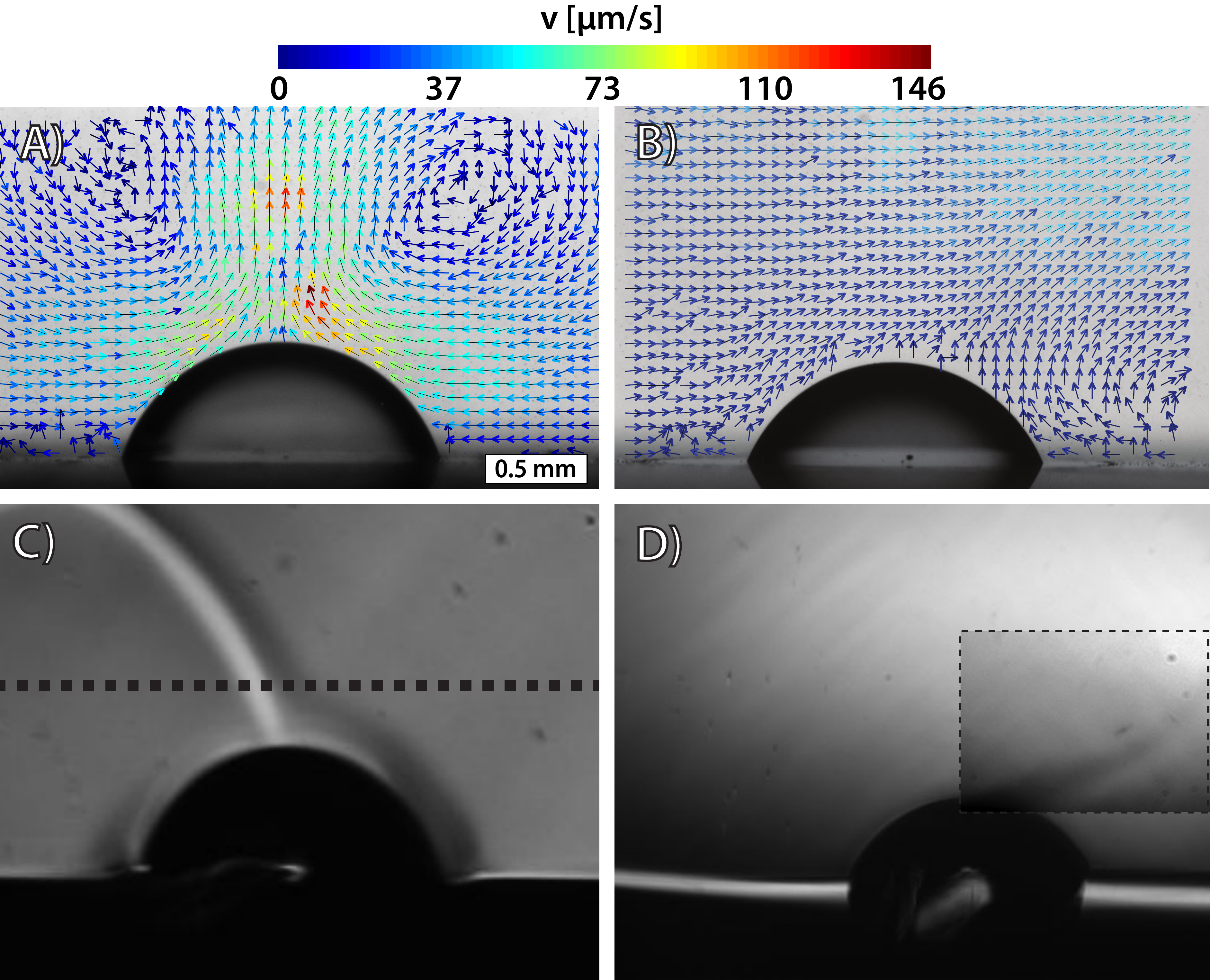}
\end{center}
\caption{\label{figPIV} (color online) Velocity fields (A\&B) and solute concentration fields (C\&D) surrounding droplets dissolving in water. A\&C show droplets of 1-pentanol, B\&D droplets of 1-heptanol. The contrast of the outlined area in D has been modified to increase the visibility of the plume. Note that panels A-C and B-D represent separate experiments. The dashed line in C indicates where the cross-sectional profiles, shown in figure \ref{figProfiles}, are taken. The images were taken approximately 30 seconds (A), 2 minutes (C) or 4 minutes (B\&D) after deposition of the droplet.}
\end{figure}

Figure 3 shows snapshots of the \si{\micro}PIV and schlieren measurements for 1-pentanol (left) and 1-heptanol droplets (right). The dissolving 1-pentanol droplet generates a clear plume originating from its apex, while fresh liquid is drawn in from the sides. The tip of the plume ends in a vortex ring which moves away from the droplet as the experiment proceeds, leaving behind a single plume (see figure \ref{figPIV}C). For 1-heptanol, with its lower solubility, the plume is far less pronounced. The particle velocities are significantly lower and the contrast of the schlieren image had to be strongly enhanced to see the plume at all (see figure \ref{figPIV}D). The weak 1-heptanol plume seems to be affected by a small mean flow in the cell, possibly caused by thermal convection due to changes in room temperature or the illumination. As we will show later on, this mean flow seems to have little influence on the dissolution behavior. Appendix A contains additional \si{\micro}PIV results, including time-resolved velocity fields around a 1-pentanol droplet, the velocity field around an insoluble sessile droplet, and the flow around a dissolving sessile droplet placed on a vertical substrate. A movie, showing the motion of the bulk around the dissolving droplet, is available online as supplementary material.

\begin{figure}
\begin{center}
\includegraphics[angle=0,width=0.5\textwidth]{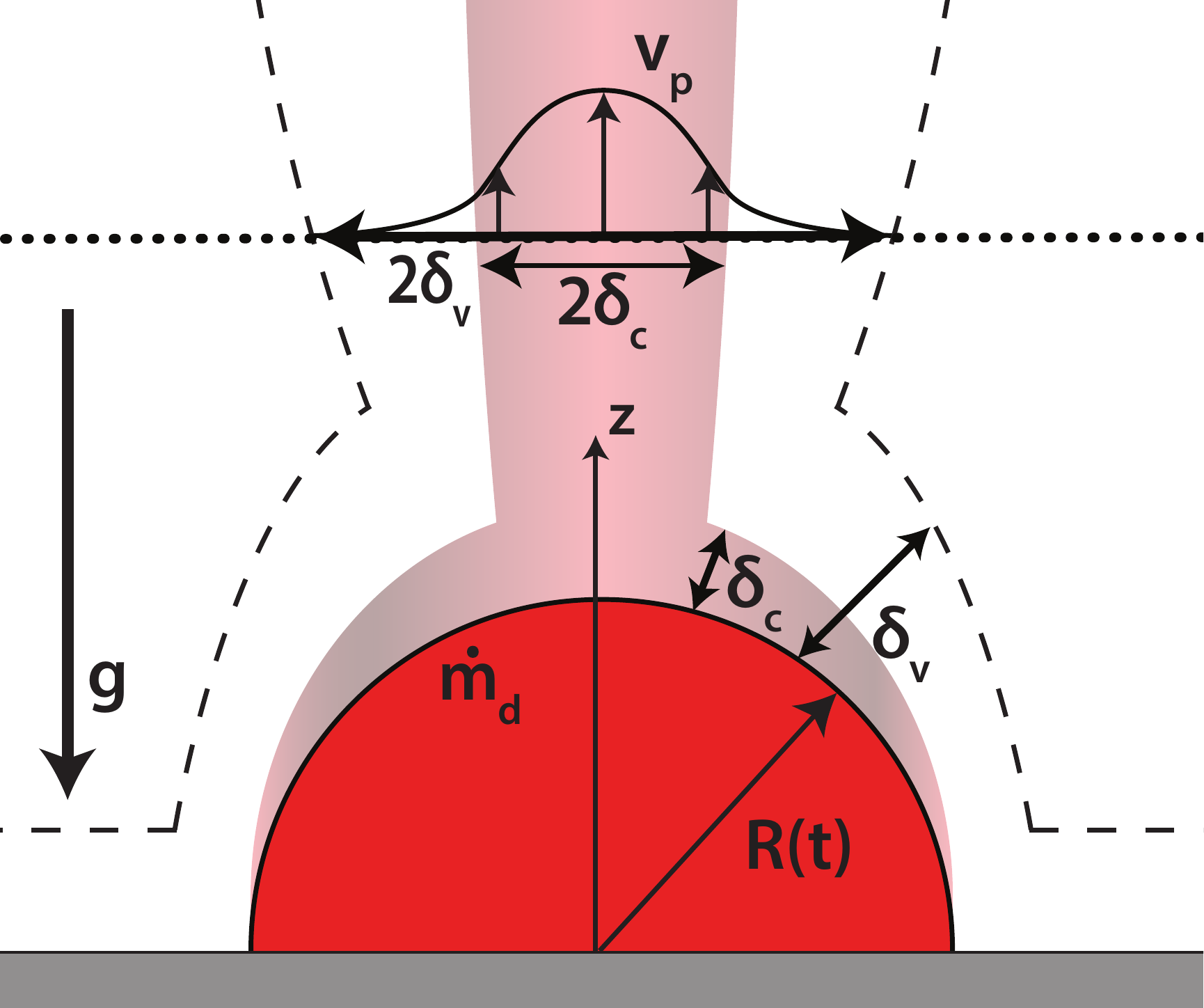}	
\end{center}
\caption{\label{figSchematic} (color online) Sketch of a sessile droplet dissolving at a rate $\dot{m}_{\text{d}}$, and convective plume. A concentration boundary layer of thickness $\delta_c$ develops over the droplet interface, since $Sc\gg1$, the velocity boundary layer has a thickness $\delta_v>\delta_c$. The boundary layers merge into plumes with diameters $2\delta_c$ and $2\delta_v$, moving at a vertical velocity $v_p$. The boundary layer widths $\delta_c$ and $\delta_v$ will change as a function of height, but are linked via the relation $\delta_v/\delta_c\sim\sqrt{\text{Sc}}$ \citep{Bejan1993}.}
\end{figure}

The two different techniques used in figure \ref{figPIV} reveal that the convective plume displays two different features, as illustrated in figure \ref{figSchematic}. Firstly, the schlieren images visualize the plume shaped region that contains dissolved alcohol. The concentration profile is characterized by a width $2\delta_c$, which increases as $\delta_c\propto \sqrt{Dz/v_p}$, with $z$ the height above the droplet, and $v_p$ the plume flow speed. Secondly, the buoyant force on the (lighter) water-alcohol mixture results in a flow, as visualized by the \si{\micro}PIV. The velocity profile of this flow (also drawn in figure \ref{figSchematic}) is characterized by a width $2\delta_v$. The liquid viscosity $\nu$ causes also the velocity profile to broaden for increasing height, namely with the same height dependence as the concentration profile, i.e. $\delta_v\sim\sqrt{\nu z/v_p}$. Therefore, the ratio $\zeta$ between the widths of the concentration and velocity profiles is fixed, $\zeta=\delta_v/\delta_c\sim \sqrt{\text{Sc}}$ \citep{Bejan1993}, where Sc is the Schmidt number $\text{Sc}\equiv\nu/D$. In the current system $\text{Sc}\approx 1200$, so $\delta_v\approx 30\delta_c$ is expected. 

\begin{figure}
\begin{center}
\includegraphics[angle=0,width=\textwidth]{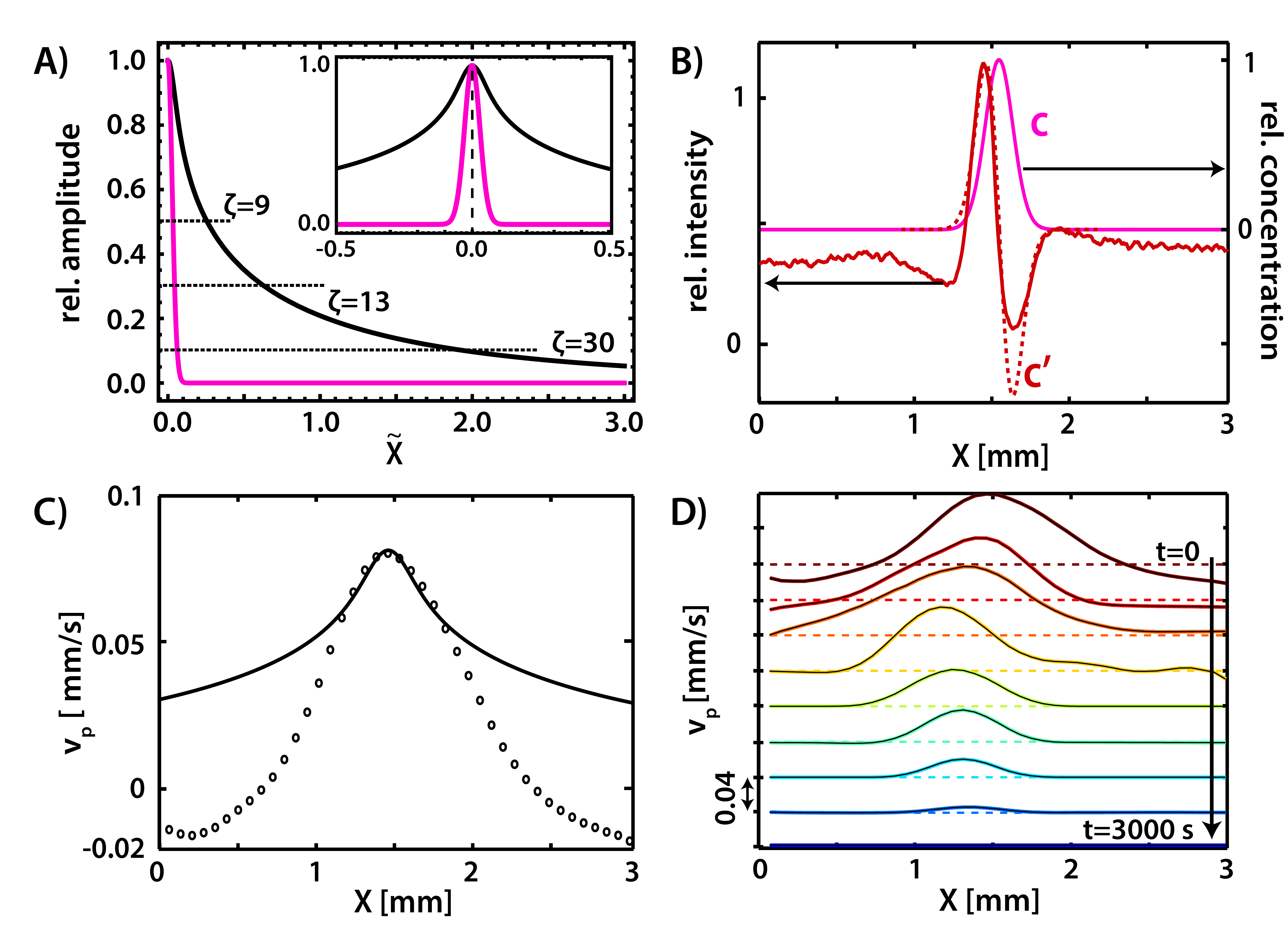}	
\end{center}
\caption{\label{figProfiles} (color online) A) Theoretical velocity profile (black) and concentration profile (pink) for a plume at Sc$=1200$. The profiles are axi-symmetric around their plume centre at dimensionless lateral coordinate $\tilde{X}=0$, as illustrated in the inset, which shows a zoom around $\tilde{X}=0$. The observed ratio $\zeta=\delta_v/\delta_c$ depends on the definition of the plume width, as illustrated by the horizontal dashed lines. The anticipated $\zeta=30$ is recovered when the plume width is evaluated at a relative amplitude of $0.1$. B) The schlieren signal (red line), measured at a horizontal cross section $300$ \si{\micro\metre} above a 1-pentanol droplet ($R_0=700$ \si{\micro\metre}), represents the derivative of the concentration profile. Therefore, both the theoretical concentration profile (pink solid curve) and its derivative (red-dashed curve) are plotted, and the derivative is fitted to match the schlieren measurement. Panel C shows the velocity profile (black circles), also measured at $300$ \si{\micro\metre} above an equally sized 1-pentanol droplet, and the theoretical velocity profile (black solid line). Panel D shows the velocity profiles at subsequent times, revealing that both the plume width and velocity decrease. For clarity, the plots are vertically shifted, with the baselines $v_p=0$ given as dashed lines. At $t=3000$ s, both droplet and plume have disappeared.}
\end{figure}

To obtain a theoretical description for the velocity and concentration profiles at Sc$=1200$, we solved equations (II.8) and (II.9), of the paper of \citet{Fujii1963}, who described the analogous case of a thermal plume above a heat source. Here, we followed the numerical procedure described by \citet{Vázquez1996}, and used equation (II.15) from Fujii's paper as a condition in the solving procedure, to obtain the velocity and concentration profiles at Sc$=1200$, as shown in figure \ref{figProfiles}A by the black and pink lines, respectively. 
The lateral coordinate $\tilde{X}$ in the theory is scaled by $\sqrt{Lz}$, with $L=\sqrt{2 \pi \nu^3/(g \beta_c \dot{m}_{\text{d}})}$, where $g$ is the acceleration of gravity, $\beta_c\equiv\frac{\partial \rho}{\partial c}/\rho_b$ is the solutal expansion coefficient, and $\dot{m}_{\text{d}}$ is the mass loss rate of the droplet. The high Sc-number in the current system results in distinct shapes for the velocity and concentration profiles. The measured value for $\zeta$ therefore depends on the definition of the plume width, as shown in \ref{figProfiles}A. The anticipated $\zeta=30$ is retrieved when evaluated at a relative amplitude of $0.1$, corresponding to the $90\%$ boundary layer definition. 

To compare our measurements to the theoretical profiles, a cross sectional intensity profile was measured in the schlieren image (figure \ref{figPIV}C) at a height of $300$ \si{\micro\metre} above the apex of the droplet (as illustrated by the dashed line in figure \ref{figPIV}C). This intensity profile is plotted as the solid red line in figure \ref{figProfiles}B, where we have to keep in mind that it represents the first derivative of the concentration profile, as it is the result from a schlieren measurement. To allow for an easy comparison, we plotted the theoretical concentration profile (pink solid line) from figure \ref{figProfiles}A and its derivative (red-dashed line) in figure \ref{figProfiles}B, and fitted the derivative to match the experiment. From this fit, a conversion factor was obtained to translate the dimensionless lateral coordinate of the theory to the length scale of the experiment. We executed the same procedure to obtain the velocity profile from the \si{\micro}PIV data in figure \ref{figPIV}A, again at $300$ \si{\micro\metre} above the droplet. This profile is plotted as the black circles in figure \ref{figProfiles}C. The theoretical velocity profile (black solid line) is superimposed on the measurement, where the previously found conversion factor was used to match the lateral coordinate. Comparison of the experimental and theoretical profiles in figure \ref{figProfiles}C reveals that while the central part of the plume shows fair agreement with the theoretical profile, the general shape of the plume is much narrower than expected from theory. The cause of this discrepancy is not understood as of yet. Possibly the vortex, substrate, and the droplet influence the plume shape, and the expected profile can be recovered when measured at higher distance above the droplet.

As mentioned before, the plume changes over time. To visualize the evolution of the plume, cross sections are taken in the \si{\micro}PIV data at subsequent times, at $300$ \si{\micro\metre} above the 1-pentanol droplet. These cross sections are plotted in figure \ref{figProfiles}D, and show that the plume properties are linked to the droplet size: both the width and the maximum velocity of the plume steadily decrease as the droplet shrinks. At $t=3000$ s, the droplet has dissolved completely.  

\section{Dissolution rate and plume velocity}\label{sec:interpretation}
The \si{\micro}PIV and schlieren images show that the convective flow around small droplets takes the form of a thin boundary layer over the droplet interface, culminating in a single plume rising from its apex. A schematic drawing of this flow and the concentration profile is shown in figure \ref{figSchematic}. If we assume (1) that at the interface of the droplet the solute concentration is constant and equal to $c_s$ and (2) that the droplet shrinks sufficiently slow. Then this situation is mathematically equivalent to the buoyant flow around a hot sphere of constant temperature $T$ and fixed radius $R$. In the context of thermal convection, the flow structure in both the boundary layer and plume are well known (see e.g. \citet{Bejan1993}, and \citet{Fujii1963}). In this section we recapitulate the main findings in terms of the dissolution problem and compare them directly to our observations.

\subsection{Dissolution rate}\label{sec:convection}

\begin{figure}
\begin{center}
\includegraphics[angle=0,width=\textwidth]{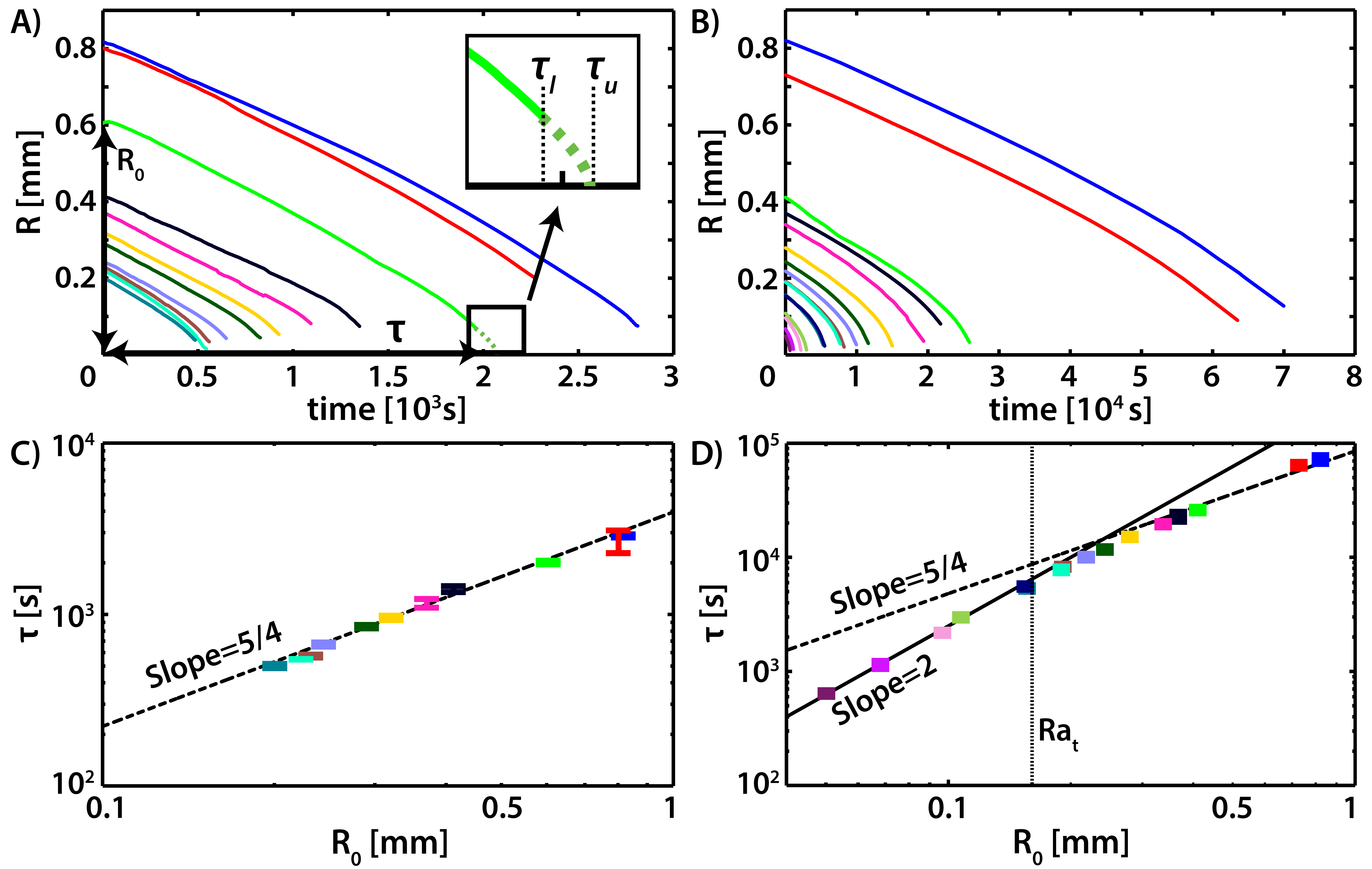}
\end{center}
\caption{\label{figRT} (color online) Equivalent radius $R=(3V/2\pi)^{1/3}$ as a function of time for 1-pentanol droplets (A) and 1-heptanol droplets (B) dissolving in water. The inset in A illustrates how for each droplet with initial radius $R_{0}$, the lifetime $\tau$ is estimated to lie between the lower estimate $\tau_\text{l}$ at the end of the experiment, and the upper estimate $\tau_\text{u}$, found by extrapolation using equation (\ref{dVdtF}) (green dashed line). The lifetime $\tau$ is plotted as a function of $R_{0}$ in figures C and D for the 1-pentanol and 1-heptanol measurements, respectively. The lines in figures C and D illustrate the $\tau\sim {R}_0^{2}$ and $\tau\sim {R}_0^{5/4}$ relations, as expected for diffusion and convection, respectively. The vertical line in (D) indicates the transition Ra-number $\text{Ra}_\text{t}=12$ which marks the transition between convection ($\text{Ra}>\text{Ra}_\text{t}$) and diffusion ($\text{Ra}<\text{Ra}_\text{t}$). For 1-pentanol, $\text{Ra}_\text{t}=12$ corresponds to $R=0.07$ \si{\milli\metre}. }
\end{figure}

\begin{figure}
\begin{center}
\includegraphics[angle=0,width=\textwidth]{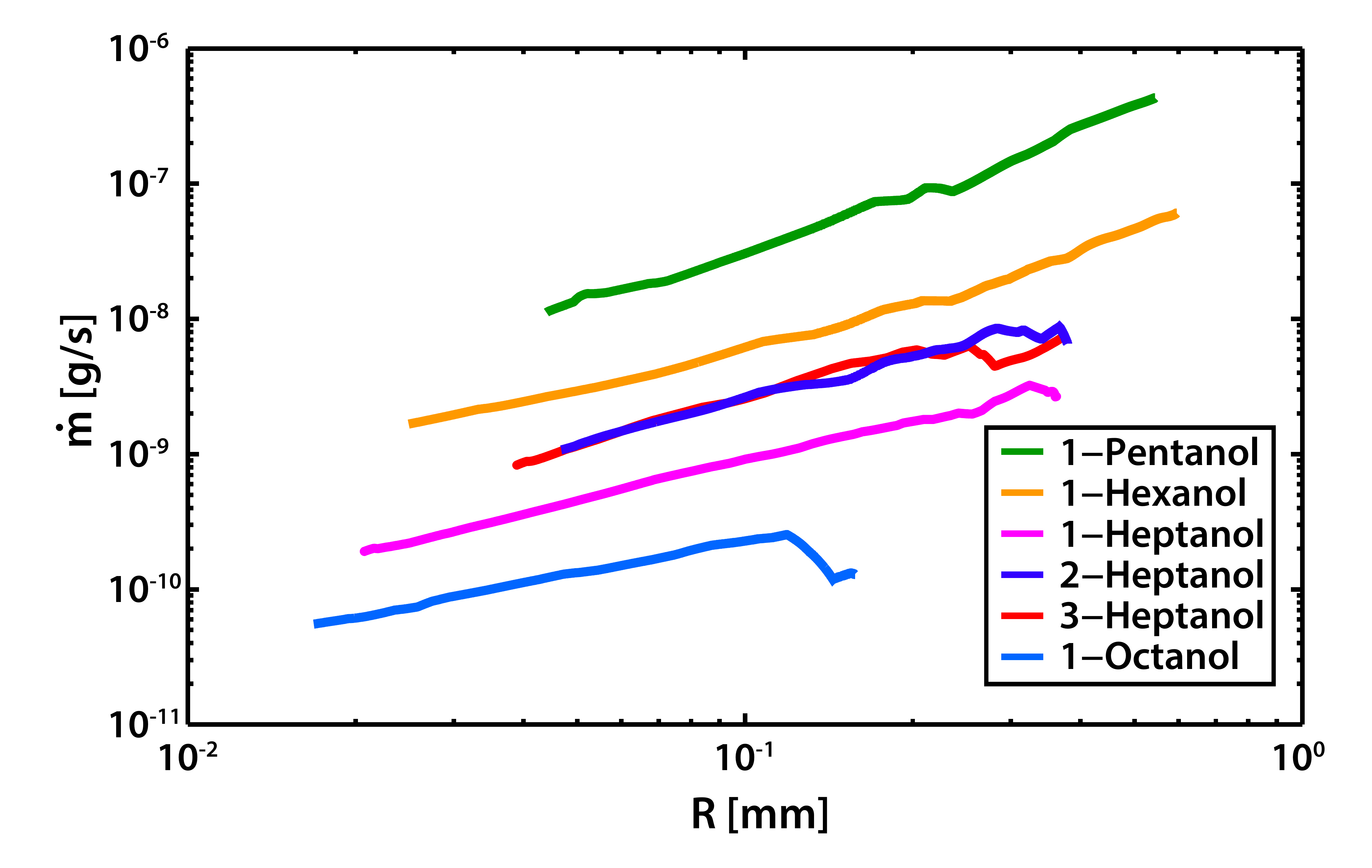}	
\end{center}
\caption{\label{dmdt} (color online) Rate of mass loss as a function of $R$ for droplets of different alcohols. The mass loss rates are ordered as a function of alcohol solubility, with 1-pentanol and 1-octanol being the best and least soluble alcohols, respectively.}
\end{figure}

To study the droplet dissolution dynamics as a function of droplet liquid and size, individual droplets of varying initial volume and alcohol type were imaged throughout the dissolution process. Since the footprint radius $R_{\text{fp}}$ shows steps due to the stick-jump mode dissolution \citep{Zhang2015, Dietrich2015,  Lohse2015}, we define the equivalent radius $R\equiv\left(3V/(2\pi)\right)^{1/3}$ to provide a continuously decreasing measure for the droplet size. Figure \ref{figRT} shows $R(t)$ for 1-pentanol (\ref{figRT}A) and 1-heptanol (\ref{figRT}B) droplets. From this, the mass loss rate $\dot{m}_{\text{d}}$ was extracted and plotted as a function of $R$ in figure \ref{dmdt} for all six alcohols. Note that while $\dot{m}<0$ for a shrinking droplet, we define the droplet mass loss rate as a positive amount, as it provides a more intuitive measure for the dissolution process. Using this, figure \ref{dmdt} shows that the measured mass loss rates for the various droplet sizes and alcohol types span two orders of magnitude. To find a universal description for the dissolution dynamics, we continue by defining dimensionless numbers which take both the droplet size and liquid into account.

\begin{figure}
\begin{center}
\includegraphics[angle=0,width=12cm]{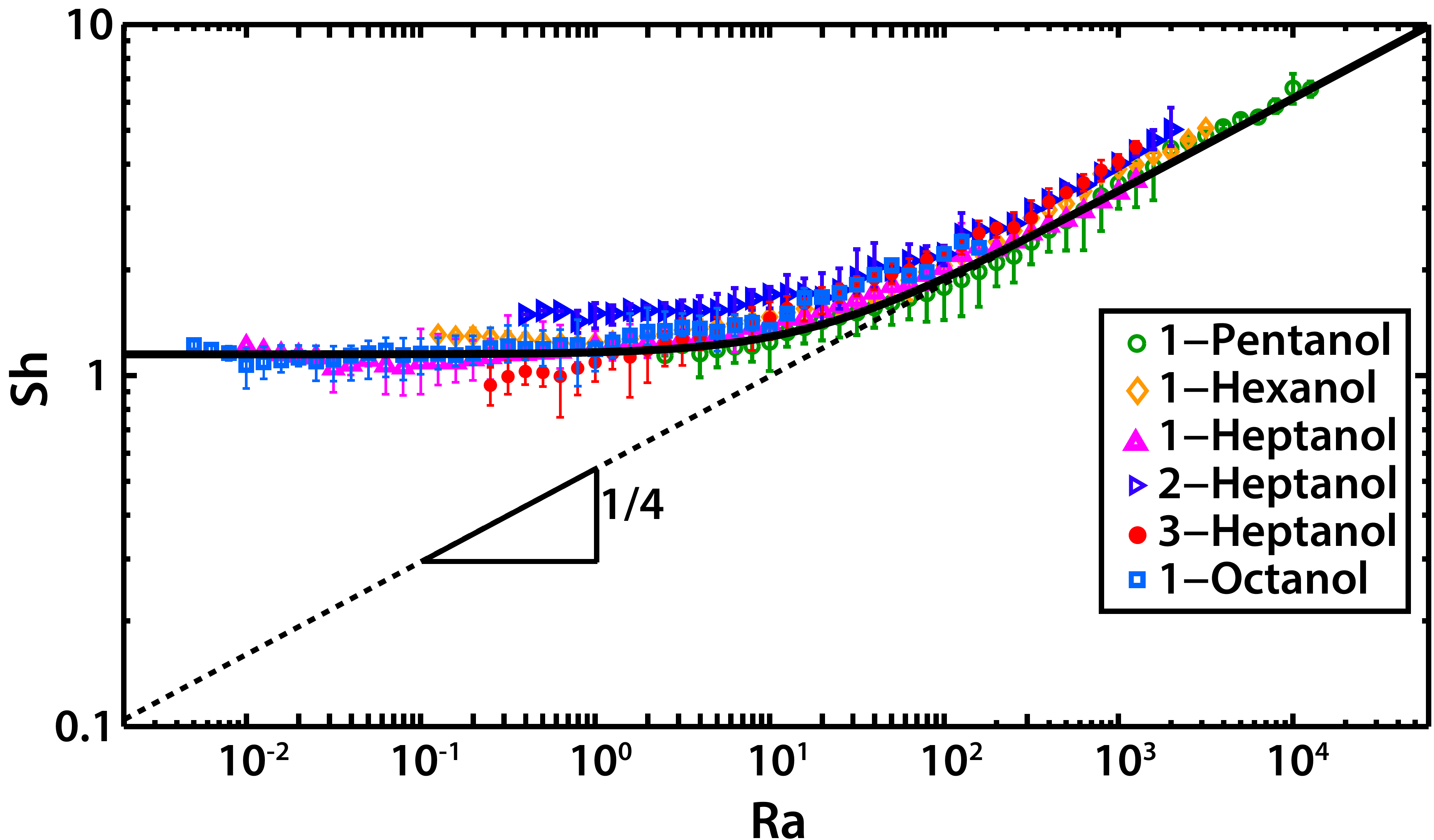}	
\end{center}
\caption{\label{figShRa} (color online) Sherwood number as a function of Rayleigh number. The plot shows the mean value and the spread for a total of $70$ measurements on droplets with initial volumes $2$ \si{\nano\litre} $\leq V_0\leq1200$ \si{\nano\litre}. Since the $\text{Ra}$-number depends both on the droplet size and its material properties, large $\text{Ra}-$number droplets are easily made using large droplets of 1-pentanol, whereas low $\text{Ra}$ droplets are best studied using small droplets of the poorly soluble 1-heptanol and 1-octanol. Equation (\ref{crossover}) is plotted as the solid black line, using $\text{Ra}_\text{t}=12.1$, $\text{Sh}_{\text{d}}=1.2$, and $n=1.0$.}
\end{figure}

In (convective) heat exchange problems, the heat exchange is usually expressed in terms of the (dimensionless) Nusselt-number, which is the ratio of the heat transfer rate and the rate for pure diffusion. The equivalent for solutal convection is the Sherwood number
\begin{equation}
\text{Sh} \equiv \frac{\mean{\dot{m_{\text{d}}}}_AR}{D\Delta c} 
\label{Sh}
\end{equation}
where $\mean{\dot{m}_{\text{d}}}_A$ is the actual (measured) mass transfer flux (rate per area), averaged over the droplet surface area $A$. This mass flux is compared to $D \Delta c/R$, the mass flux of pure (steady) diffusion from an equally sized spherical droplet (or a sessile droplet with $\theta = 90^{\circ}$). In the case of pure diffusion from our sessile droplets ($45^{\circ}<\theta<75^{\circ}$), we expect to find a diffusion-limited Sherwood number $0.9<\text{Sh}_{\text{d}}<1.3$, of which the exact value depends on the droplet contact angle, as discussed in appendix B.
For the case of laminar flow at high Sc-number, \citet{Bejan1993} provides a complete and insightful derivation of the momentum equation, showing that the flow can be described using the Boussinesq approximation of the Navier-Stokes equation. This approximation assumes a slender boundary layer (i.e., $\delta_c\ll R$), a constant pressure over the width of the boundary layer, and a limited density difference. For high $\text{Sc}$ numbers, the buoyant force is balanced by viscosity and it can be shown that $\delta_c/R\sim \text{Ra}^{-1/4}$, independent of Sc \citep{Bejan1993}. Here $\text{Ra}$ is the Rayleigh number which is the ratio of the buoyant force to the damping force
\begin{equation}
\text{Ra} \equiv \frac{g\beta_c\Delta c {R}^3}{\nu D} 
\label{Ra}
\end{equation}
Taking $\delta_c$ as the typical length scale over which diffusion takes place in the presence of convection, we find $\mean{\dot{m}_{\text{d}}}_A \sim D\Delta c/\delta_c$, so that
\begin{equation}
\text{Sh}\sim R/\delta_c\sim \text{Ra}^{1/4},
\label{ShRa}
\end{equation}
again independent of Sc. If we recast the data from figure \ref{dmdt} in terms of the $\text{Ra}$ and $\text{Sh}$-numbers, all data sets from the six different alcohols collapse, as shown in figure \ref{figShRa}. This figure also reveals that for large $\text{Ra}$, the data follow the anticipated $\text{Sh}\sim \text{Ra}^{1/4}$ scaling, which is plotted as the dashed line. For small $\text{Ra}$ numbers, $\text{Sh}$ converges to a plateau, as expected for diffusion. It is noteworthy that the $\text{Sh}\left(\text{Ra}\right)$ dependence from \citet{Enriquez2014}, who studied the growth of CO$_2$ gas bubbles in slightly supersaturated water, is almost identical to our figure \ref{figShRa}. This indicates that the flow structures around bubbles and droplets are very similar.

To better understand the transition between the convective and the diffusive behavior, and to find the value of the transition Ra-number $\text{Ra}_{\text{t}}$, we fit a crossover function of the form 
\begin{equation}
\text{Sh}(\text{Ra}) = \text{Sh}_{\text{d}}\left(1+\left(\frac{\text{Ra}}{\text{Ra}_{\text{t}}}\right)^n\right)^{\frac{1}{4n}}
\label{crossover}
\end{equation} 
to the data. Here $n$ is a fitting parameter which describes the sharpness of the transition. Equation (\ref{crossover}) was fitted to the individual datasets of each alcohol, to obtain $\text{Ra}_{\text{t}}=12.1\pm 5.8$, $\text{Sh}_{\text{d}}=1.2\pm0.2$, and $n=1.0\pm0.5$. Equation (\ref{crossover}) is plotted in figure \ref{figShRa}, using the mean values. The fitted curve confirms that for $\text{Ra}>\text{Ra}_{\text{t}}\approx12$, the data follows the $\text{Sh}\sim\text{Ra}^{1/4}$ scaling. A transition exists around $\text{Ra}_{\text{t}}$, where the contribution of convective mass transport gradually decreases. When $\text{Ra}<\text{Ra}_{\text{t}}$, we obtain the diffusive limit $\text{Sh}\approx1.2$, independent of $\text{Ra}$.

\subsection{Plume velocity}

Similar to the concentration boundary layer around the droplet, the local velocity and structure of a convective plume are determined by a competition between buoyant and viscous stresses. For a thermal plume, this is described by the \emph{local} thermal Rayleigh number $\text{Ra}{(z)}$, based on the local temperature difference between the center of the plume and the surroundings (see e.g. \citet{Fujii1963} and \citet{Vázquez1996}). The local solutal Rayleigh number can be written similarly, based on the local concentration difference $\Delta c(z)$:
\begin{equation}
\text{Ra}(z) \equiv \frac{g\beta_c\Delta c(z) z^3}{\nu D} 
\label{Raz}
\end{equation}
The $\text{Ra}(z)$-number can be conveniently written in terms of $\dot{m}_{\text{d}}$ by using $\Delta c \sim \dot{m}_{\text{d}}/(Dz)$ \citep{Fujii1963}:
\begin{equation}
\text{Ra}(z)\sim\frac{g\beta_c\dot{m}_{\text{d}}z^2}{D^2\nu}
\end{equation}
 Again by exploiting the analogy with the thermal case, one finds the following scaling behaviors for the width of the plume $\delta_{c(z)}$ and the central velocity $v_p$ of a solutal plume with $\text{Sc}\gg1$ \citep{Fujii1963}:
 \begin{equation}
 \delta_c\sim z\left(\text{Ra}(z)\right)^{-1/4}\propto z^{1/2}
 \end{equation}
 \begin{equation}
 v_p\sim\frac{D}{z}\left(\text{Ra}(z)\right)^{1/2}=\left(\frac{g\beta_c \dot{m}_{\text{d}}}{\nu}\right)^{1/2}
\label{Vp}
 \end{equation}
 Note that $v_p$ is independent of $z$. We use the droplet dimension $R$ to non-dimensionalize the plume velocity, and define a plume Reynolds number:
 \begin{equation}
 \text{Re}_p\equiv\frac{v_p{R}}{\nu}
\label{Re}
 \end{equation}
 By using the relation for $v_p$ from (\ref{Vp}) we obtain
 \begin{equation}
 \text{Re}_p\sim\left(\frac{g\beta_c\dot{m}_{\text{d}}{R}^2}{\nu^3}\right)^{1/2},
 \end{equation}
in which we can insert the previously found expression for $\dot{m}_{\text{d}}$ to obtain
 \begin{equation}
 \text{Re}_p\sim\ \text{Ra}^{5/8}\text{Sc}^{-1}
\label{ReScale}
 \end{equation}

\begin{figure}
\begin{center}
\includegraphics[angle=0,width=12cm]{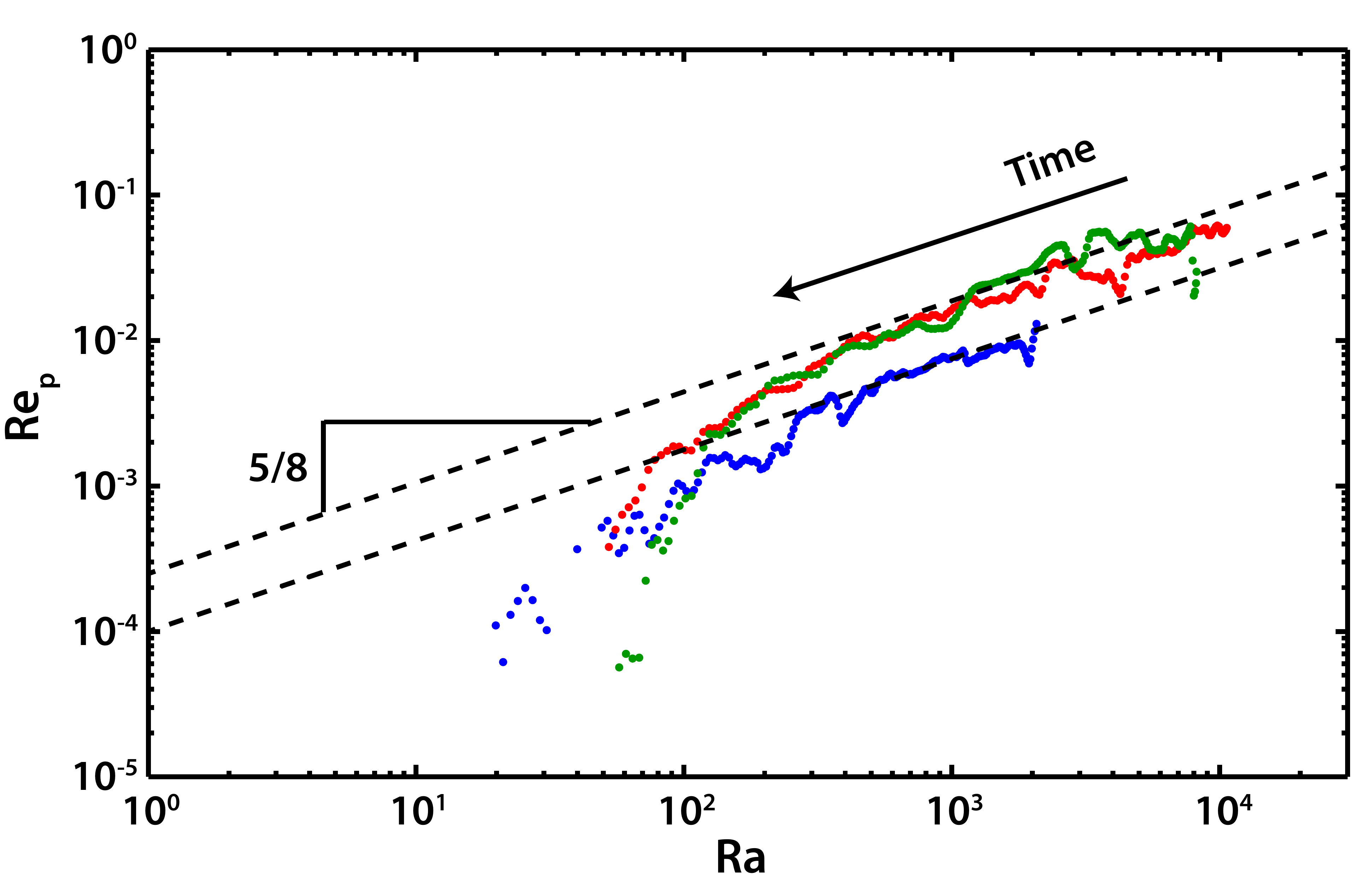}	
\end{center}
\caption{\label{figReRa} (color online) The plume Reynolds number as a function of $\text{Ra}$-number for three individual 1-pentanol droplets with $V_0=700$ nl (red), $V_0=550$ nl (green), and $V_0=140$ nl (blue). The expected $\text{Re}\sim \text{Ra}^{5/8}\text{Sc}^{-1}$ scaling is indeed found when $\text{Ra}\gg\text{Ra}_{\text{t}}$, but breaks down when Ra approaches $\text{Ra}_{\text{t}}$. This marks the onset of the transition region (located around $\text{Ra}_{\text{t}}$), in which convective transport is gradually exceeded by diffusive transport. To obtain the fits to the two measurements, plotted in red and blue an additional pre-factor of $0.25$ (red/green) and $0.1$ (blue) was required.}
\end{figure}

To test this scaling, we measure the maximum vertical velocity in the \si{\micro}PIV data at a height of $300$ \si{\micro\metre} above the droplet, together with the size of the droplet and use this to calculate $\text{Re}_p(\text{Ra})$. The result of this analysis is shown in figure \ref{figReRa}. For $\text{Ra}\gg\text{Ra}_{\text{t}}$, we find the anticipated $\text{Re}_p\sim \text{Ra}^{5/8}$. However, when $\text{Ra}$ approaches $\text{Ra}_{\text{t}}$, a transition from convection to diffusion occurs. During and below this transition, equation (\ref{ReScale}) does not hold anymore, and the plume rapidly fades away. Note that this transition occurs when still $\text{Ra}>\text{Ra}_{\text{t}}$. For the measurements shown, there seems to be some dependence of the scaling pre-factor on the initial size of the droplet. The smallest droplet displays a somewhat lower overall plume velocity than the two larger ones. We are not yet able to pinpoint the reason for this difference. One possible hypothesis might be that inertia of the bulk flow plays a role, and that the larger droplets create a stronger convection that persists throughout the dissolution. However, more work is required to confirm or rebut this hypothesis.

\section{Dissolution time}
The convective flow and the related increase in mass transport was described in the previous section. In this section we proceed by deriving an expression for the convective droplet dissolution rate and associated dissolution time $\tau_{\text{c}}$. However, we start by briefly introducing diffusive dissolution, which we will use later on. 

An expression for the diffusive volume loss rate $dV/dt$ has been given by \citet{Popov2005} in the context of evaporating sessile water droplets. Popov's solution can be rewritten to find the rate of change of the droplet, expressed in terms of the previously introduced equivalent radius $R$
\begin{equation}
\frac{dR}{dt} = -\frac{D\Delta c}{2 \rho R}f(\theta)\left[\frac{2}{2-3\cos\theta+\cos^3\theta}\right]^{1/3}\sin{\theta}
\label{dVdtF}
\end{equation}
where 
\begin{equation}
f(\theta) = \frac{\sin(\theta)}{1+\cos(\theta)}+4\int_0^\infty \frac{1+\cosh(2\theta\epsilon)}{\sinh(2\pi\epsilon)}\tanh[(\pi-\theta)\epsilon]\mathrm{d}\epsilon
\end{equation}
is a geometrical shape factor to describe the effect of the impenetrable substrate. Note that for simplicity we have neglected the intermittent contact line pinning which was observed in the experiments \citep{Zhang2015, Dietrich2015}, and equation (\ref{dVdtF}) describes dissolution in the constant contact angle mode. Integration of equation (\ref{dVdtF}) results in the dissolution time, with the associated diffusive timescale given by equation (\ref{difftime}), i.e. in particular $\tau_{\text{d}}\propto R_0^2$.

We can perform a similar calculation for the convective mass exchange, again based on the cooling sphere analogy. An important difference between the cooling sphere and our dissolving droplet is that in the latter case, the radius decreases in time. However, if the dissolution is slow we can assume the process to be quasi-static and neglect this effect. We start by equating the rate of mass loss, $\dot{m}_{\text{d}}\sim -R^2\rho(dR/dt)>0$, to the rate at which mass is carried away in the convective plume $\dot{m}_p =AD\Delta c \text{Sh}/R\sim AD\Delta c \text{Ra}^{1/4}$/R, with $A\propto R^2$ the area of the droplet-bulk interface. From this we obtain:
\begin{equation}
\frac{dR}{dt}=-a\left(\frac{g\beta_c \Delta c_s^5D^3}{\nu \rho_{\text{d}}^4 {R}}\right)^{1/4}
\label{convection}
\end{equation}
with a prefactor $a$ of order 1. Separation of variables and integrating $R$ from $R=R_0$ till $R=0$, and time from $t=0$ till $t=\tau_{\text{c}}$, gives the dissolution time with the associated convective time scale $\tau_{\text{c}}$ with
\begin{equation}
\tau_{\text{c}}=\left(\frac{\nu \rho_{\text{d}}^4 R^5_0}{g\beta_c\Delta c^5D^3}\right)^{1/4}.
\label{tauc}
\end{equation}
Therefore, for droplets dissolving in the convection dominated regime, we expect a dissolution time $\tau_{\text{c}} \propto {R}_0 ^{5/4}$, with a material-dependent prefactor.

To test this scaling behavior, we used the $R(t)$ curves in figures \ref{figRT}A and B and extracted the dissolution time from each droplet. Since the droplets could not be measured until complete dissolution, the actual value of $\tau$ had to be estimated. Therefore, we assumed that the last stage of the dissolution process was diffusion limited and extrapolated the droplet evolution by integrating equation (\ref{dVdtF}), using the smallest still measured droplet size of the experiment as initial value. This extrapolation (illustrated by the green dotted line in the inset of figure \ref{figRT}A) provides the upper bound of $\tau$, whereas the lower bound is given by the time at which the experiment was terminated. The thus obtained values for $\tau$ are plotted as a function of $R_0$ for 1-pentanol and 1-heptanol in figures \ref{figRT}C and \ref{figRT}D, respectively. For larger droplets we indeed find $\tau\propto {R}_0^{5/4}$ as expected from equation (\ref{tauc}), while for smaller 1-heptanol droplets $\tau\propto {R}_{0}^{2}$ is found, as expected for pure diffusion. The vertical line in figure \ref{figRT}D indicates $\text{Ra}=12$, showing that the transition from diffusive to convective dissolution occurs around $\text{Ra}_{\text{t}}$, consistent with our findings in section 4.

\begin{figure}
\begin{center}
\includegraphics[angle=0,width=\textwidth]{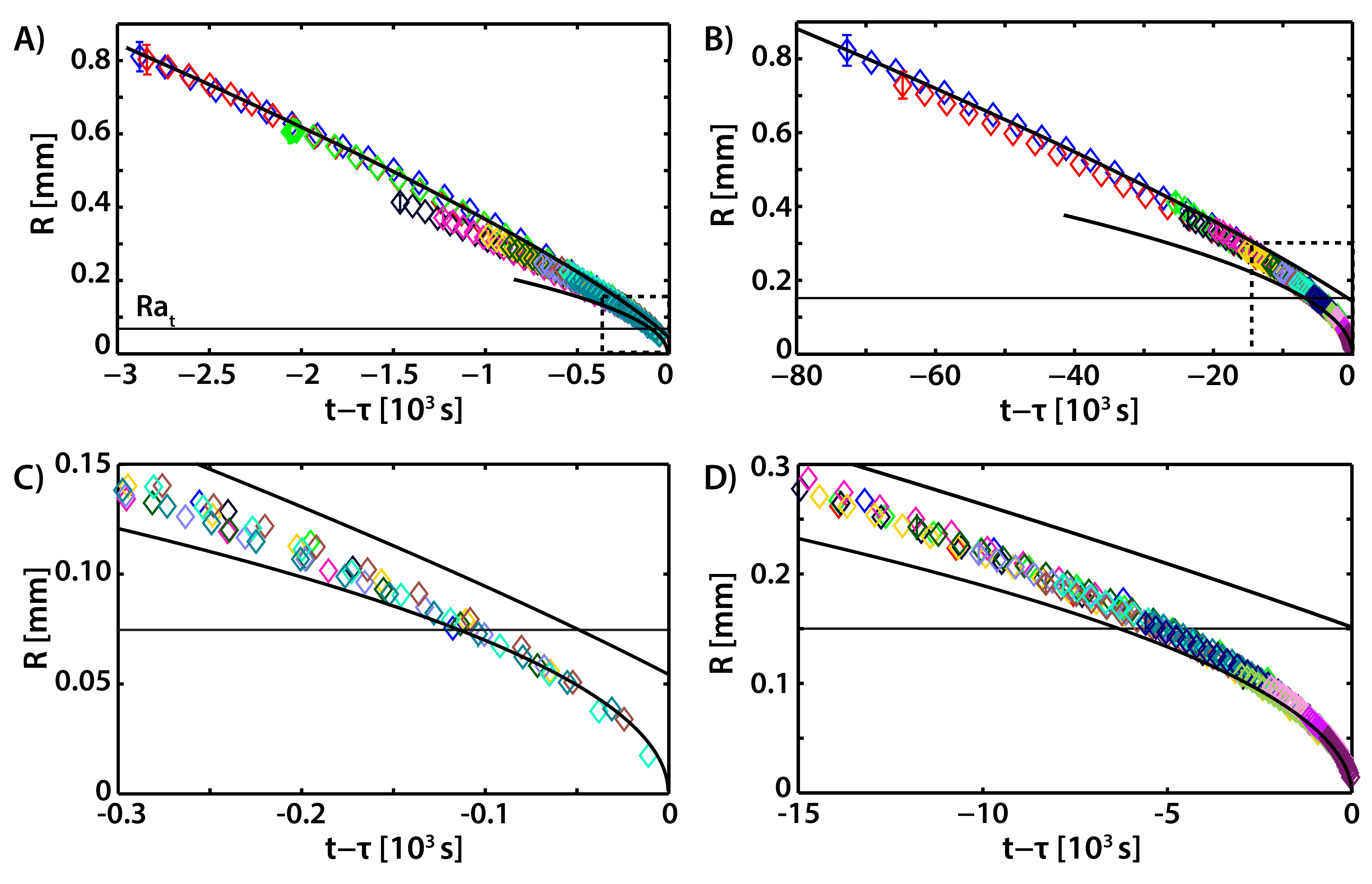}
\end{center}
\caption{\label{detail} (color online) Equivalent radius $R=(3V/2\pi)^{1/3}$ as a function of time to complete dissolution $t-\tau$ for 1-pentanol droplets (A) and 1-heptanol droplets (B). Panels C and D provide a zoom of the outlined parts of A and B, respectively. The upper and lower black curves in each panels represents the integration of equations (\ref{convection}) and (\ref{dVdtF}), respectively. When $\text{Ra}>\text{Ra}_{\text{t}}=12$, equation (\ref{convection}) accurately describes the dissolution dynamics. Around $\text{Ra}_{\text{t}}$, which is indicated by the horizontal line in each panel, a transition exists, in which convection is progressively overtaken by diffusion. In the final stage ($\text{Ra}<\text{Ra}_{\text{t}}$), the dissolution is well described by diffusion, i.e., equation (\ref{dVdtF}). The color coding for the individual measurements is the same as in figure \ref{figRT}.}
\end{figure}

Now that we have confirmed the $\tau_{\text{c}}\propto R_0^{5/4}$ behavior for large droplets, we finally test whether equations (\ref{dVdtF}) and (\ref{convection}) provide accurate descriptions of the dissolution dynamics in the diffusive and convective regimes, respectively. Moreover, we can test whether the transition between these regimes indeed occurs around $\text{Ra}_\text{t}=12$, as found before. To this purpose, the curves in figures \ref{figRT}A and \ref{figRT}B are replotted in figures \ref{detail}A and \ref{detail}B as a function of the time $t-\tau$. Figures \ref{detail}C and \ref{detail}D provide a close-up of the final stage of dissolution, the outlined parts of panels A and B, respectively. Along with the experimental data, we plotted the numerical integration of of the convective dissolution model, equation (\ref{convection}), which is the upper black line in all figures. We also plotted the diffusive dissolution model, equation (\ref{dVdtF}), represented by the lower black line in each figure. All figures show that for $\text{Ra}\gg\text{Ra}_{\text{t}}$, equation (\ref{convection}) accurately captures the droplet dissolution process, reflecting convection-dominated dissolution. The value for the prefactor $a$ in equation (\ref{convection}) was adjusted for each alcohol to obtain a good fit in the convective regime. The values used are listed in table \ref{table2}, along with $\theta$, for all alcohols. 
For $\text{Ra}\approx\text{Ra}_{\text{t}}$, the overlap between our convective model and the experiments becomes worse, consistent with the transition from convection-dominated to diffusion-dominated dissolution.
The final stage of dissolution is best observed in figures \ref{detail}C and \ref{detail}D. When $\text{Ra}<\text{Ra}_{\text{t}}$, the dissolution is purely diffusive, reflected by the good overlap between equation (\ref{dVdtF}) and the measurements. The above findings confirm the applicability of our convective dissolution model when $\text{Ra}>\text{Ra}_{\text{t}}$, and validate $\text{Ra}_{\text{t}}\approx12$ as the transitional Ra-number for the transition from convective to diffusive dissolution dynamics. 

\begin{table}
\begin{centering}
\caption{Value of the dimensionless prefactor $a$ in equation (\ref{convection}) for all alcohol types as determined from a fit to the data, together with the observed contact angles of these droplets on the PFDTS-coated substrates in water.}
\begin{tabular}{ |c | c | c | c | c |c|c|}
\hline
Alcohol 	&  \emph{1-pentanol}		& \emph{1-hexanol}	& \emph{1-heptanol}	&\emph{2-heptanol}	&\emph{3-heptanol}	&\emph{1-octanol}	\\
\hline
   		&					&				&				&				&				&				\\
$a$		& $0.65 \pm0.05$			&  $0.65 \pm0.05$		&  $0.55 \pm0.05$		& $0.65 \pm0.05$		& $0.7 \pm0.05$		& $0.65 \pm0.05$	\\
$\theta$	&$70^{\circ}$			& $70^{\circ}$		& $70^{\circ}$		& $52^{\circ}$		&$45^{\circ}$		& $70^{\circ}$\\
\hline
  \end{tabular}
\label{table2}
\end{centering}
\end{table}

\section{Conclusion}

Sessile droplets of long-chain alcohols immersed in water are ideally suited to experimentally study the basic laws of mass transfer around small objects. By choosing alcohols of different chain lengths, the alcohol's solubility in water can be varied by almost two orders of magnitude, while its other properties remain practically the same. This large range of solubilities allowed us to vary the convective driving parameter, the Rayleigh-number, by over six orders of magnitude, while keeping the droplets large enough to visualize their shrinkage and the flow around them.

Using a combination of \si{\micro}PIV and schlieren technique, we directly demonstrated that above a transition Rayleigh number $\text{Ra}_{\text{t}}\approx12$, a buoyant flow develops around a dissolving droplet, due to the density differences between the lighter alcohol-water mixture, as compared to the heavier clean water. By modeling the observed boundary layer structure at the droplet interface and in the plume, we derived a basic scaling relation $\text{Sh}\sim \text{Ra}^{1/4}$ for convective mass transport. Using this relation as a starting point, we derived expressions for the shrinkage rate of the droplet and the velocity of the plume. In the convective regime, these models are in good agreement with our data. However, once the droplet dissolves to a size close to $\text{Ra}_{\text{t}}$, diffusion gradually overtakes convective mass transport and the convective scaling relations break down.

The observed convection and associated increase in mass transport confirms earlier work on growing bubbles in supersaturated water \citep{Enriquez2014,Penas2015}, indicating that it is a universal phenomenon.

The value for $\text{Ra}_{\text{t}}$ presented in this work provides an important indication to determine the dominant transport mechanism in droplet dissolution. In conjunction with this, the convective dissolution model allows for more accurate predictions of droplet dissolution times. The demonstrated predictability of the dissolution behavior of single, sessile droplets on a horizontal substrate, invites to test the derived scaling relations and fitting parameters in more complicated situations. For example, the applicability of the derived scaling relations and measured value for Ra$_{\text{t}}$ could be tested in the context of bubble growth or droplet evaporation. Moreover, interesting changes in the flow profile can be expected, for example, when the orientation or the wettability of the substrate is changed. Other possible research directions include placing multiple droplets close together, to study their interaction, or making the droplets so large ($R>\lambda_c$) that they form a puddle.


\section*{Appendix A: \si{\micro}PIV Data}
\begin{figure}
\begin{center}
\includegraphics[angle=0,width=\textwidth]{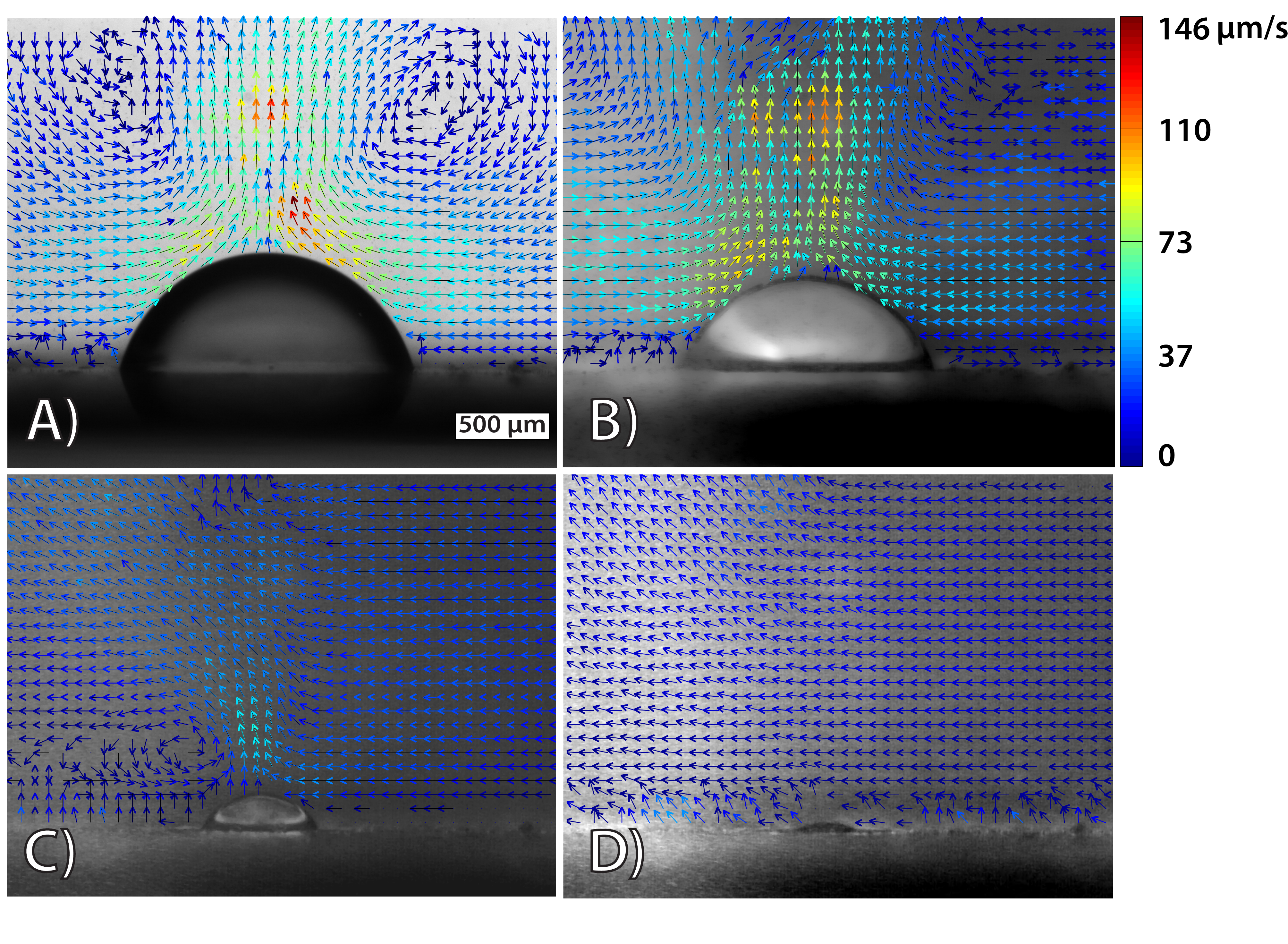}
\end{center}
\caption{\label{figure1app} Velocity fields in water around a dissolving 1-pentanol droplet, 30 s after deposition of the droplet (A), and $600$ s (B), $2000$ s (C), and $2800$ s (D) after deposition.}
\end{figure}

The flow structure around the dissolving droplet has been measured using micro particle image velocimetry (\si{\micro}PIV), utilizing the procedure and setup described above. In figure \ref{figure1app}, the time evolution of the velocity field around a dissolving 1-pentanol droplet ($V_0=700$ nl) is shown. Shortly after deposition of the droplet (figure \ref{figure1}A), the data reveals the presence of a toroidal vortex above the droplet, and a strong, narrow plume originating from the droplet apex. At $600$ \si{\second} into the dissolution process, shown in figure \ref{figure1app}B, the center of the vortex is observed to have moved away from the droplet, causing the plume to slow down and broaden. Figure \ref{figure1app}C shows the droplet at $2000$ \si{\second} after deposition: the droplet has shrunk considerably, however, a small plume is still visible, along with a small mean flow, from right to left. Close to the end of the experiment ($t=2800$ \si{\second}, panel D), the droplet has almost disappeared, as has the plume. A small right-to-left mean flow is still observable.

\begin{figure}
\begin{center}
\includegraphics[angle=0,width=12cm]{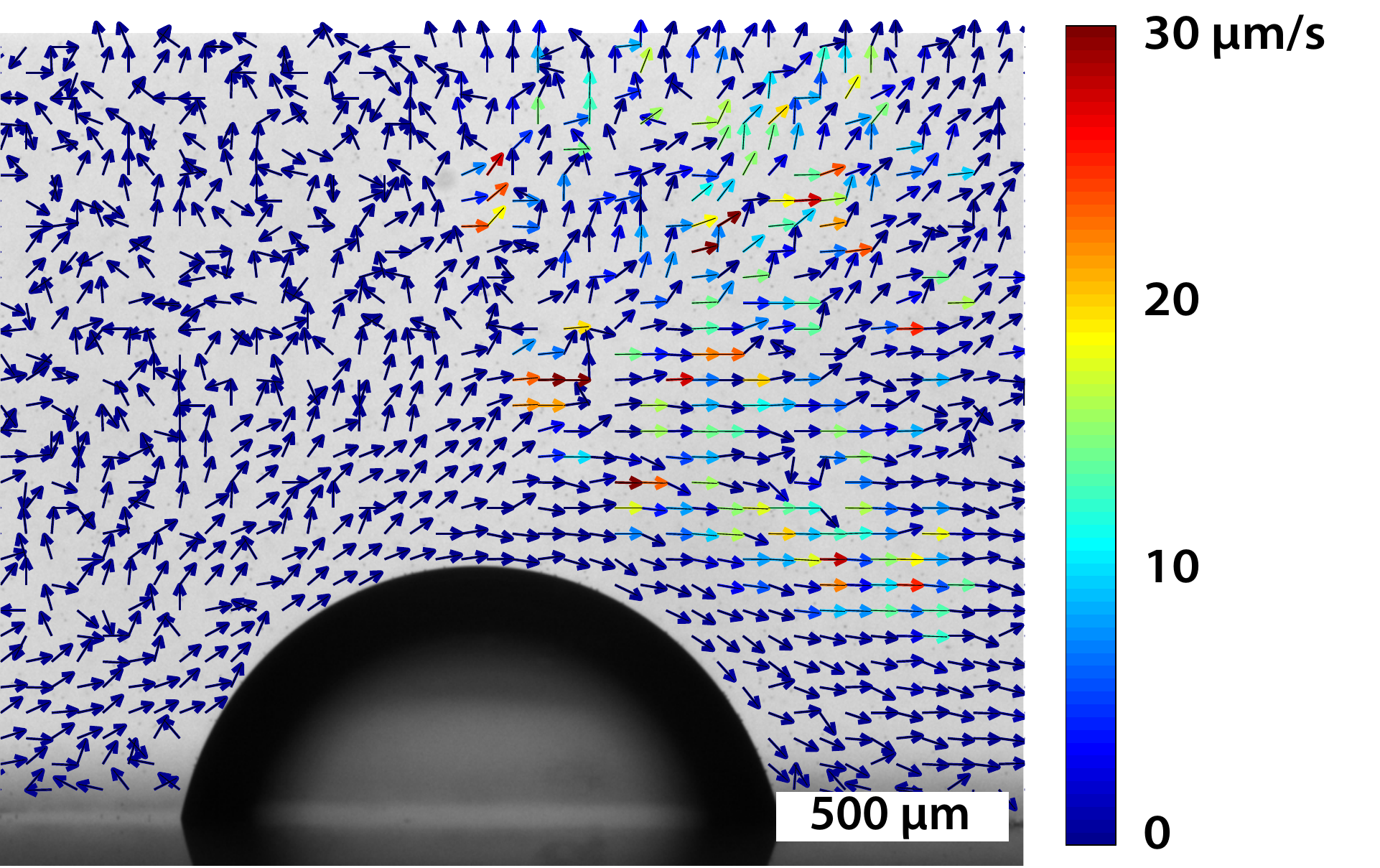}
\end{center}
\caption{\label{figure2app} Velocity fields in water around an in-soluble 1-decanol droplet.}
\end{figure}

To show that the plume is caused by solutal convection, and not simply by the presence of a spherical object at the interface, the experiment is repeated using an equally sized droplet of 1-decanol. This alcohol has negligible solubility ($c_s=0.037$ \si{\gram\per\litre} \citep{Kinoshita1958}). For this droplet, $Ra\approx10$ which means that solutal convection should be absent. The velocity field around this 1-decanol droplet is shown in figure \ref{figure2app}. From this figure, it is clear that the presence of the droplet does not cause the formation of a plume. A slight mean flow is present, which can be seen to flow \emph{around} the droplet. Although invisible from figure \ref{figure2}, it should be noted that in this particular experiment, tracer particles adhered to the droplet interface, something that did not happen in all other experiments where soluble droplets were studied.

\begin{figure}
\begin{center}
\includegraphics[angle=0,width=12cm]{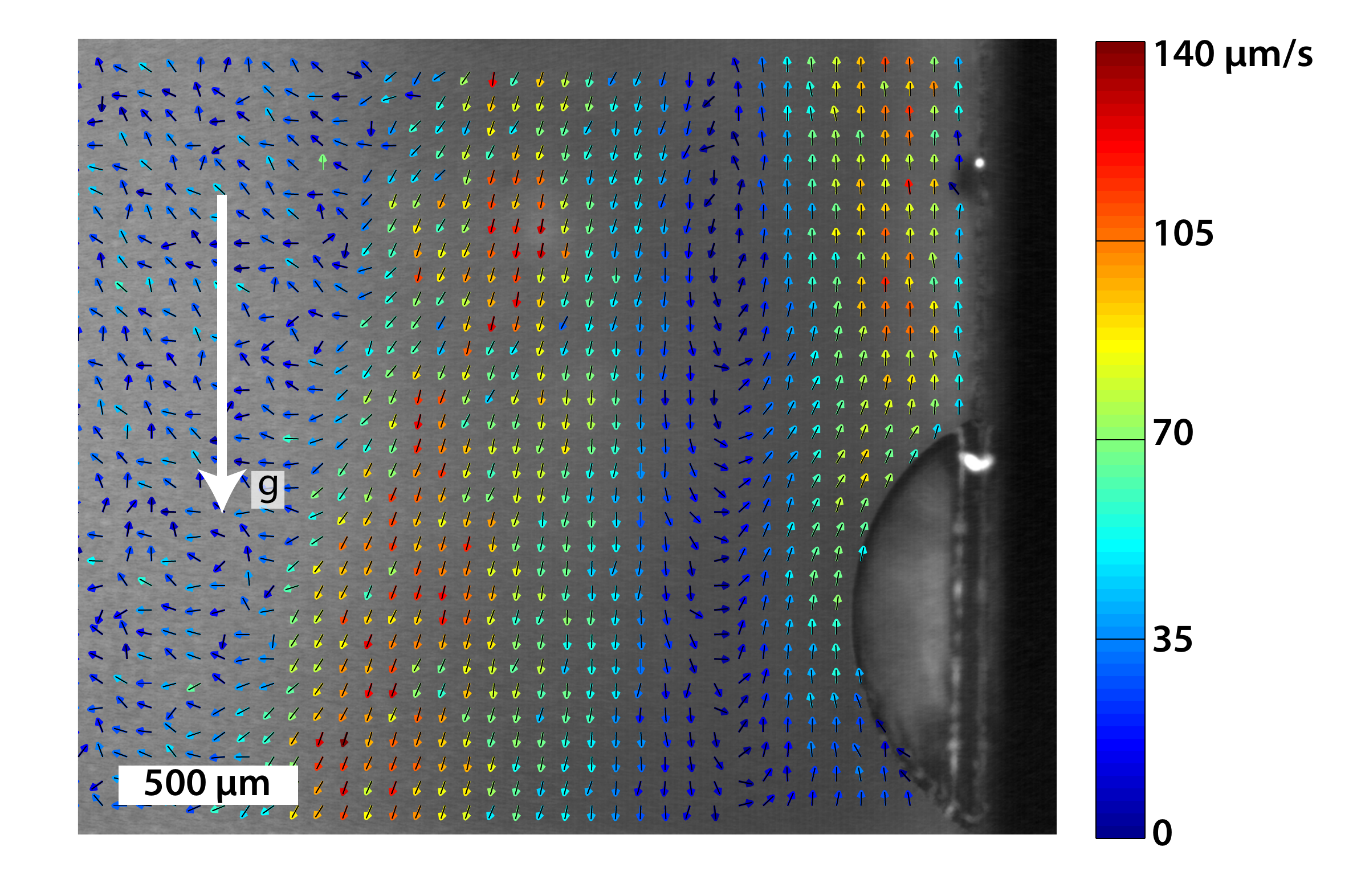}
\end{center}
\caption{\label{figure3} Velocity fields in water around a dissolving 1-pentanol droplet, sitting on a vertical wall, measured 300 s after deposition.}
\end{figure}

Hypothetically, the convection could be induced by surface tension gradients as well, developing over the droplet-water interface. These gradients would cause Marangoni convection over the alcohol-water interface \citep{Kostarev2004}. If this would be the case, the plume would always have the same shape and orientation with respect to the droplet and substrate, regardless the direction of gravity. To check whether this is the case, the substrate with a 1-pentanol droplet in place is mounted vertically in the center of the tank. The resulting flow, shown in figure \ref{figure3}, is found to be mainly parallel to the substrate. Liquid is replenished by inflow from the side and bottom, creating a large convection roll. The fact that the plume orients in a direction opposite to gravity, regardless the orientation of the substrate, confirms that the convection is buoyancy driven.

\section*{Appendix B: Derivation of the Sherwood number in the diffusion limited case}

The Sherwood number has been defined in equation (\ref{Sh}). In the diffusion limited case, the mass loss rate $\dot{m}$ can be calculated from the droplet properties and its size. For a spherical droplet with radius $R$, floating in an infinite bulk, the steady state mass loss is \citep{Epstein1950}
\begin{equation}
\frac{dm}{dt}=-4\pi R D \Delta c, 
\label{dMdt}
\end{equation}
resulting in a Sherwood number of $\text{Sh}_{\text{d}}=1$.

\begin{figure}
\begin{center}
\includegraphics[angle=0,width=12cm]{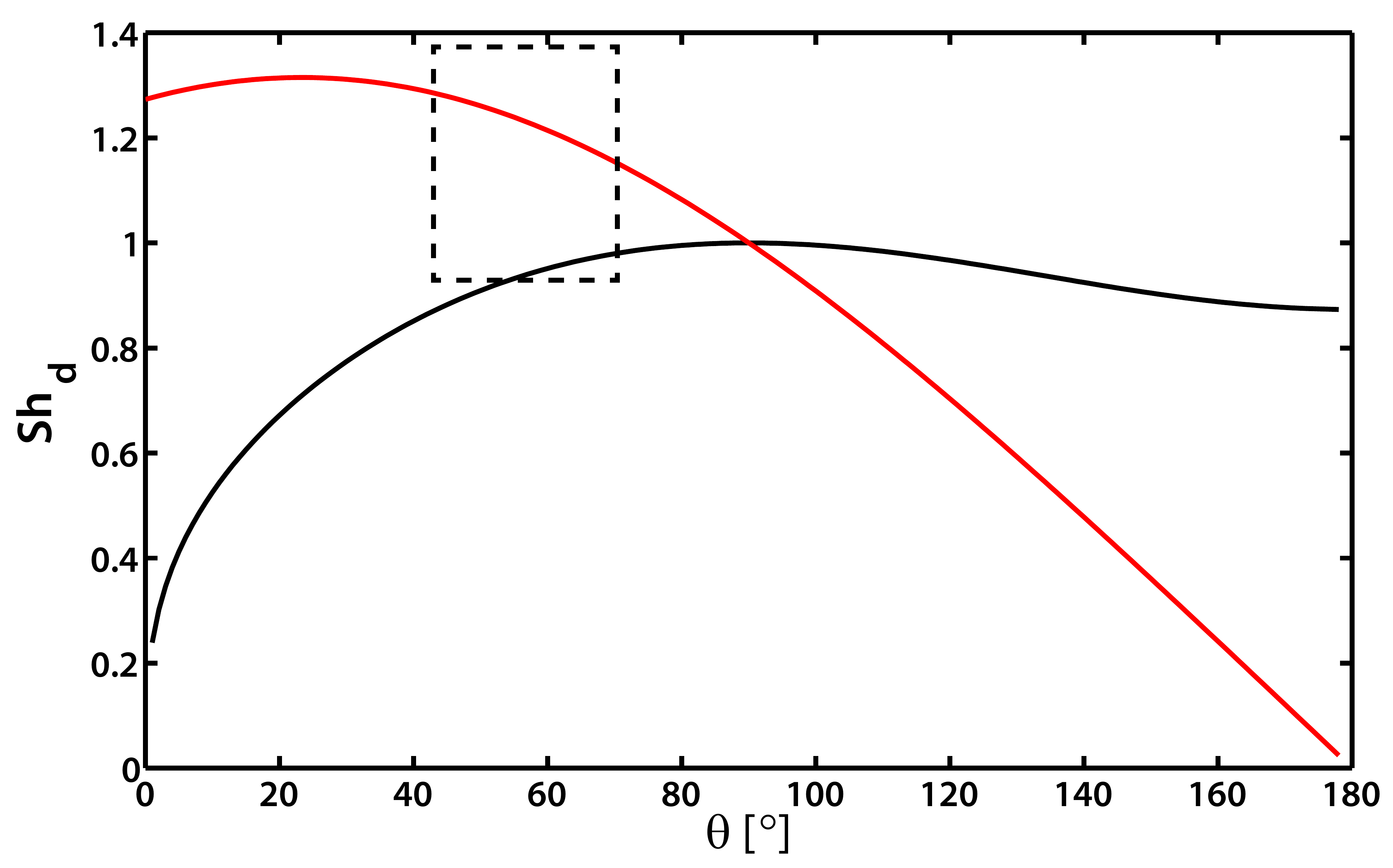}
\end{center}
\caption{\label{figure6app} Diffusion limited Sherwood number $\text{Sh}_\text{d}$, as a function of the droplet contact angle $\theta$, calculated using the equivalent radius $R$ (black curve) or the footprint radius $R_\text{fp}$ (red curve), as characteristic length scale. Both curves cross at $\theta=90^{\circ}$, corresponding to a hemisphere with $R_\text{fp}=R$ and Sh$_\text{d}=1$. The dotted region indicates the range of contact angles and experimentally observed values for $\text{Sh}_\text{d}$.}
\end{figure}

For a sessile droplet, the presence of a substrate changes the dissolution, and a suitable correction factor has to be used \citep{Popov2005}:
 \begin{equation}
\frac{dm}{dt} =-\pi R_{\rm{\rm{fp}}} D\Delta cf(\theta)
\label{dVdtFapp}
\end{equation}
with
\begin{equation}
f(\theta) = \frac{\sin(\theta)}{1+\cos(\theta)}+4\int_0^\infty \frac{1+\cosh(2\theta\epsilon)}{\sinh(2\pi\epsilon)}\tanh[(\pi-\theta)\epsilon]\mathrm{d}\epsilon
\end{equation}
and $R_{\rm{\rm{fp}}}$ the footprint diameter of the droplet. Using goniometry, both $R_{\rm{\rm{fp}}}$ and the droplet surface area $A$ can be expressed in terms of the volume (and thus the equivalent radius $R=(3V/2\pi)^{1/3}$), and the contact angle. From this, the Sherwood number for a sessile droplet with contact angle $\theta$, dissolving purely via diffusion is found to be
 \begin{equation}
\text{Sh}_{\text{d}} =\frac{f(\theta)}    {\sin\theta\left(1+\tan^2\frac{\theta}{2}\right)\left(\frac{2}{2-3\cos\theta+\cos^3\theta}\right)^{1/3}},
\label{ShD}
\end{equation}
which indeed only depends on the droplet contact angle. Note that we defined the mass loss rate as a positive quantity, and dropped the minus sign from equation (\ref{dVdtFapp}), resulting Sh$_{\text{d}}>0$. By solving $f(\theta)$ numerically, $\text{Sh}_{\text{d}}(\theta)$ is plotted as the black line in figure \ref{figure6app}. When $\theta=90$\si{\degree}, the droplet has the shape of a hemisphere, and $\text{Sh}_{\text{d}}$ is equal to that of a free sphere, $\text{Sh}_{\text{d}}=1$. When $\theta>90$\si{\degree} the mass transport is reduced (as compared to that from a free sphere) and hence $\text{Sh}_{\text{d}}<1$. 

When $\theta<90$\si{\degree}, $\text{Sh}_{\text{d}}$ decreases towards zero, which is an implication of the choice of our characteristic length scale: For practical reasons, the equivalent radius $R$ is chosen as the characteristic length scale. In the extreme case of dissolution from a flat disk, $\theta\rightarrow 0$, $V\rightarrow0$, and thus $R\rightarrow0$ resulting in $\text{Sh}_{\text{d}}=0$. This does not provide a proper physical representation of the actual system, as it would result in zero mass exchange in the case of complete wetting.

Then an alternative characteristic length scale is the footprint radius $R_{\rm{\rm{fp}}}$, in which case $\text{Sh}_\text{d}$ is given by the red curve in figure \ref{figure6app}. By using $R_{\rm{\rm{fp}}}$ as the characteristic length scale, $\text{Sh}_\text{d}$ for evaporation from a flat disk with radius $R_{\rm{fp}}$, can be calculated exactly. $f(\theta=0)=4/\pi$ \citep{Stauber2014}, which gives $\text{Sh}_\text{d}=\frac{4}{\pi}$. However, the drawback of using $R_{\rm{\rm{fp}}}$ as the length scale appears for $\theta>90$\si{\degree}. Especially when $\theta\rightarrow180^{\circ}$, $R_{\rm{\rm{fp}}}\rightarrow0$, resulting once again in $\text{Sh}_\text{d}\rightarrow0$. 

So what experimental value for $\text{Sh}_{\text{d}}$ is expected? The Sherwood number scales mass exchange with respect to a diffusive, free, and spherical droplet. Hence $\text{Sh}_{\text{d}}=1$ when $\theta=90$\si{\degree}. Due to the substrate, the mass loss from a surface droplet with $\theta<90$\si{\degree} is \emph{larger} as compared to the mass exchange from the same segment of a free and spherical droplet \citep{Hu2002}. The opposite is true when $\theta>90^{\circ}$ \citep{Stauber2015}. Still, $\text{Sh}_{\text{d}}$ is always of order $1$ for practical droplets ($10^{\circ}<\theta<160^{\circ}$), and independent of droplet size. The detailed dependence, as proposed in figure \ref{figure6app} could be the subject of future work, where small droplets (ensuring $\text{Ra}<10$) dissolve on substrates of various wettability. 

\section*{Acknowledegments}
We gratefully acknowledge funding from The Netherlands Organisation for Scientific Research (NWO, project number 11431) and from the NWO Spinoza programme.

\bibliographystyle{jfm}
\bibliography{PIVBib}
\end{document}